\documentclass[epsf]{article}
\usepackage{epsf,amsfonts,amssymb}
\usepackage{amssymb}
\usepackage{amsmath}

\begin{document}

\title{Analysis of Data Clusters Obtained by Self-Organizing Methods}
\author{V.V. Gafiychuk* \thanks{Institute Applied Problems of Mechanics and
Mathematics}\thanks{Institute of Computer Modeling, Cracow
University of Technology, Cracow, Poland}, B.Yo. Datsko*, J.
Izmaylova* }\date{\today} \maketitle
\begin{abstract}
The self-organizing methods were used for the investigation of
financial market. As an example we consider data time-series of
Dow Jones index for the years 2002-2003 (R. Mantegna,
cond-mat/9802256). In order to reveal new structures in stock
market behavior of the companies drawing up Dow Jones index we
apply SOM (Self-Organizing Maps) and GMDH (Group Method of Data
Handling) algorithms. Using SOM techniques we obtain SOM-maps that
establish a new relationship in market structure. Analysis of the
obtained clusters was made by GMDH.
 \\
{}\\
\noindent \textbf{keywords}: self-organizing map, group method of
data handling.
\end{abstract}

\section{Introduction}

Complex systems we meet in economics and finances are usually
described by huge amounts of data, parameters and often have a
non-predictable and chaotic behavior. To reveal interrelations
among these parameters and data new methods of computer analysis
have to be developed. These problems are of exceptional interest
now when computer technologies provide enormous possibilities for
collecting, storing and processing of information obtained by
tracing system behavior. It is believed that this information is
particularly important for the prediction of the system behavior.
In this case researchers dealing with such types of systems often
try to apply statistical \cite{princ,cox} or neurons methods
\cite{ref,koh} for the discovery of new knowledge about the
system. In framework of this approach self-adjustable methods
known as self-organization methods \cite{ref,koh,4,1,3} are
especially interesting because they could be applied in
autonomous regime without external setting for every particular
case. In this article we use two types of self-organization
methods for investigation of financial market behavior by
analyzing data series of Dow Jones index. The study of the
financial markets behavior is a very important and actual question
because financial markets are described by inadequate prior
information; they include many variables and are characterized by
fuzzy elements. Therefore, market structures are extremely
difficult for analytical description, and their dynamics is hard
to predict.

\section{Processing data by self-organizing methods}

We investigate data time-series of Dow Jones index for the years
2002-2003\footnote{AA-Alcoa, ALD-Allied Capital, AXP-American
Express Co, BA-Boeing Co, BS-Bethlehem Steel, CAT-Caterpillar
Inc., CVX-Chevron Corp.\& Texaco Inc., DD - Du Pont, DIS-Walt
Disney Co., EK-Eastman Kodak Co., GE-General Electric, GM-General
Motors, GT-Goodyear Tire, IBM-IBM Corp., IP-International Paper,
JPM-Morgan JP, KO-Coca Cola Co., MCD-McDonalds Corp.,
MMM-Minnesota Mining, MO-Philips Morris, MRK-Merck\& Co Inc.,
PG-Procter\& Gamble, S-Sears Roebuck, T-AT\& T, UN-Unilever,
UTX-United Tech, XOM-Exxon Corp}. These data are presented in a
table, each column of which contains information about
corresponding stock price during each working day for the years
2002 and 2003 separately. Each line contains information about
every day state of all indices. We use time-series of the prices
for 27 companies drawing up Dow Jones index, which are calculated
at the end of each working day.\footnote{For more details see
http://finance.yahoo.com}

\begin{table}[h]
\epsfxsize=2.3in {\epsfbox[80 400 350 500]{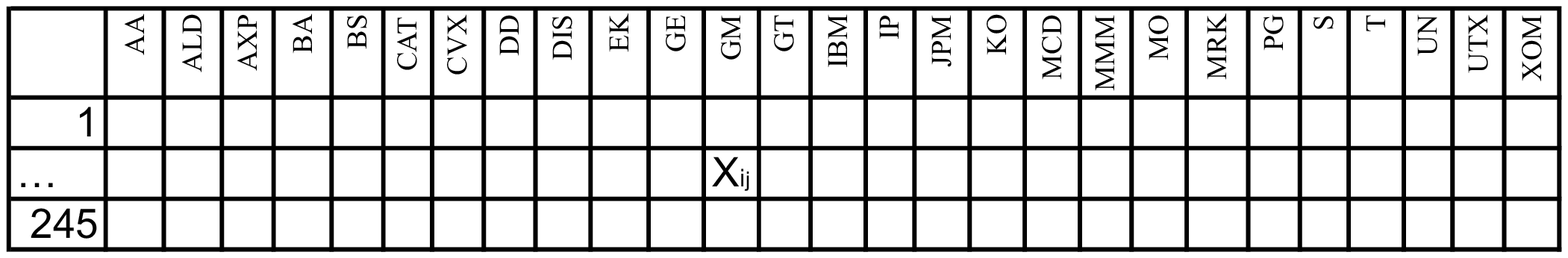}}
\caption{Input data}\label{data}
\end{table}

For finding relationships in collective dynamics of prices we
have normalized the input data

\begin{equation}\label{(norm)}
  x_{ji}=\frac{x'_{ji}-x^{min}_{i}}{x^{max}_{i}-x^{min}_{i}},\quad
j=\overline{1,245},i=\overline{1,27},
\end{equation}
where $x'_{ji}$ is non-normalized elements of input Table 1,
$x^{max}_{i}$ and $x^{min}_{i}$ are maximal and minimal
non-normalized elements of $i$-column, respectively. In order to
reveal some structures in stock market behavior for the companies
under consideration we apply SOM (Self-Organizing Maps) \cite{4,5}
and GMDH (Group Method of Data Handling) algorithms \cite{1,7}
and compare these results. \par The SOM is an
unsupervised-learning neural-network method that provides a
similarity graph of input data. It represents obtained results in
a very convenient way. A typical simplified version of the SOM
algorithm consists of two steps iterated for every sample
\cite{4}: a)  finding the best matching units; b) adaptation of
the weights. Initially all neuron weights are initialized by
uniform distribution on the interval $[0,1]$. The distance
between two neurons of the two-dimensional grid is found as

\begin{equation}\label{eucl}
  d_j=\|x-w_j\|=\sqrt{\sum_{i=0}^{N-1}(x_i-w_{ij})^2},
\end{equation}
where $j$ is the index of neuron in the net, $i$ is the dummy
index of vector components, $w_{ij}$ is the weight of synapse,
which matches $i$ - component of input vector with output neuron
$j$. Thus, we find for each vector $x$ such a neuron $c$, for
which the distance between $x$ and $c$ is the smallest

\begin{equation}\label{eucl1}
c\equiv\|x-w_c\|=\min_j d_j.
\end{equation}
Vectors of weights $w_j$ are adapted using the following rule

\begin{equation}\label{eucl12}
w_j(t+1)=\left\{\begin{array}{l}w_j(t)+\alpha(t)h_{cj}(t)\cdot|x(t)-w_j(t)|,
j\in N_c,\\ w_j(t),\quad j\notin N_c\\
\end{array}\right.
\end{equation}
where $N_c$ describes the neighborhood of ''neuron-winner''
(\ref{eucl1}), $\alpha(t)$ is the learning-rate factor, which
decreases monotonously with the regression steps, $h_{cj}(t)$ is
a scalar multiplier called the neighborhood function. Thus,
training process leads to reduction of the distance between input
signal and position of ''neuron-winner'' as well as to the
reduction of the Euclidean distance between input vector and any
vector $w_j,\; j\in N_c$. The SOM-map is obtained as the result of
this mapping of vector x on neurons plane. Similar input vectors
are placed closely to each other on the SOM-map. Such procedure
makes it possible to single out the input information and locate
the input vectors in the vicinity of similar vectors on the map
without preliminary training, using only internal properties of
input data. Next processing of the input data by chosen rule
(\ref{eucl12}) leads to the neuron-grids training and formation of
corresponding clusters. In this case the mapping of the
information by the system of neurons is characterized by the
adaptation of system to the information. Since the similar data
are placed in the same clusters the introduction of a new vector
that is similar to existing ones does not change clusters
distribution significantly. But input of data vector, which have
not been presented on the map until yet, leads to the rearranging
of the mapping structure and formation of new clusters.
Optimization of clusters and formation of SOM-grid are carried
out in the process of training and are based on the procedure of
minimization of errors squared sum in existing clusters. Applying
the above algorithm to the input data from Table 1 for the year
$2002$ ($27$ input vectors of $N$-dimension, $N=245$) we obtain
SOM-map and SOM-grid presented on Fig.1 and Fig. 2.

\begin{figure}[tbph]
\begin{center} \epsfxsize=1.5in {\epsfbox[170 10 400
470]{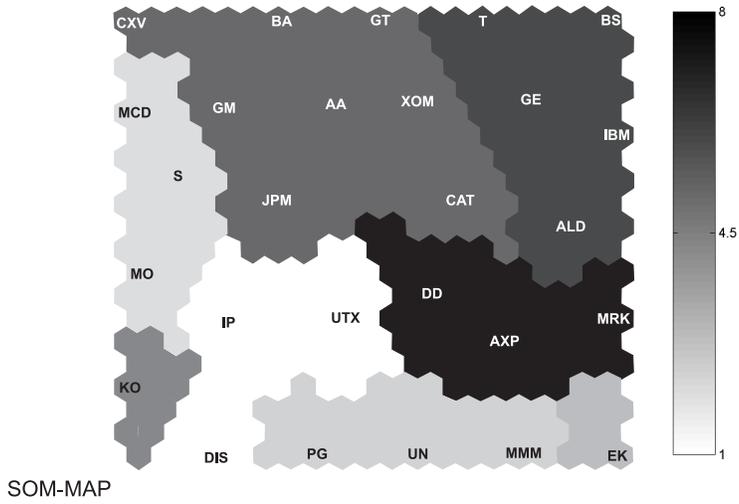}} \caption{SOM-map 20x20. Data taken for the
year 2002.} \label{map}
\end{center}
\end{figure}

\begin{figure}[tbph]
\begin{center} \epsfxsize=2in {\epsfbox[100 200 350 600]
{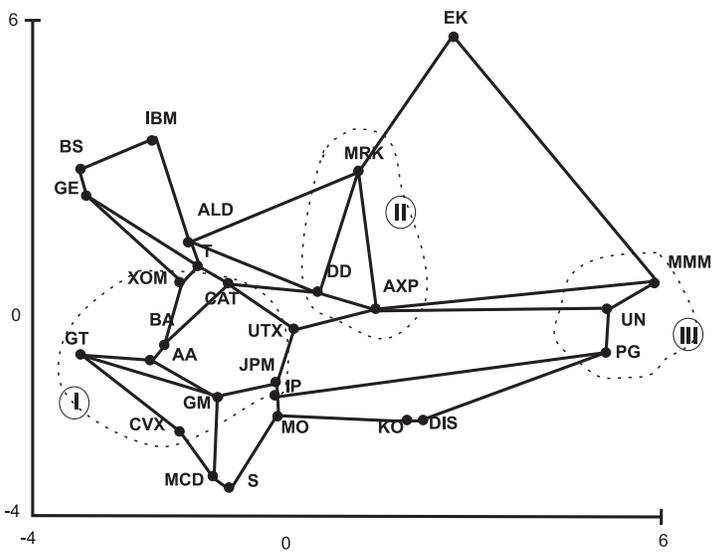}} \caption{SOM-grid. Data taken for the year 2002.}
\label{norm}
\end{center}
\end{figure}

Let us consider economic interpretation of the SOM-map and
SOM-grid obtained for data presented in the Table 1. Elements of
the first cluster I are companies associated with transport
branch (BA, GM, CAT - machine building, GT - technical branch,
CVX, GM, XOM - chemical industry associated with production of
oil and fuel). At the same time EK and PG, also engaged in
chemical industry, are placed in the separate cluster far from
other chemical companies. This fact shows their difference from
all other companies. EK is basically engaged in chemical
technologies associated with film-production, polymeric materials
etc., that makes it dependent on mining companies (MRK, MMM). The
second company (PG) is a world-famous manufacturer of goods for
self- and household care.
\par Analyzing the second cluster II we can see that it consists of companies
involved in health- and life-care manufacturing. DD develops new
technologies of nutrition production and MRK is a global
research-driven pharmaceutical company. The last company of this
cluster (AXP) is engaged in financial and international banking
services worldwide. Considering KO and DIS companies one can
notice that in spite of the fact that they are located in separate
clusters their proximity shows similar dynamics of stock prices.
\par The elements of the third cluster III (UN, PG, MMM) compose a
subsystem, which provide goods for household care and personal
products. One can understand the relationship between PG
(mentioned above) and UN, which is also engaged in production of
goods for self- and household care. So, these companies are
presented in health care and safety markets.
\par It should be noted that SOM-map and SOM-grid are not changed
when they are smoothed over input data by the method of the
nearest neighborhood and by Savitsky-Golay method up to the range
M=10 \cite{2}. Thus SOM-grids preserve their topology on such
time intervals. It indicates that the stock market also preserves
its collective properties on time intervals which are shorter
then 10 days. This fact can be used for short term forecasts
during share price oscillation in the market.

Processing data time-series of Dow Jones index for the year 2003
with SOM-algorithm, we obtain the results, which are a little bit
different from those for 2002 (Fig.3). In fact one can notice
that EK and MRK are separated from other indices.

\begin{figure}[tbph]
\epsfxsize=2.3in {\epsfbox[100 50 400 490]{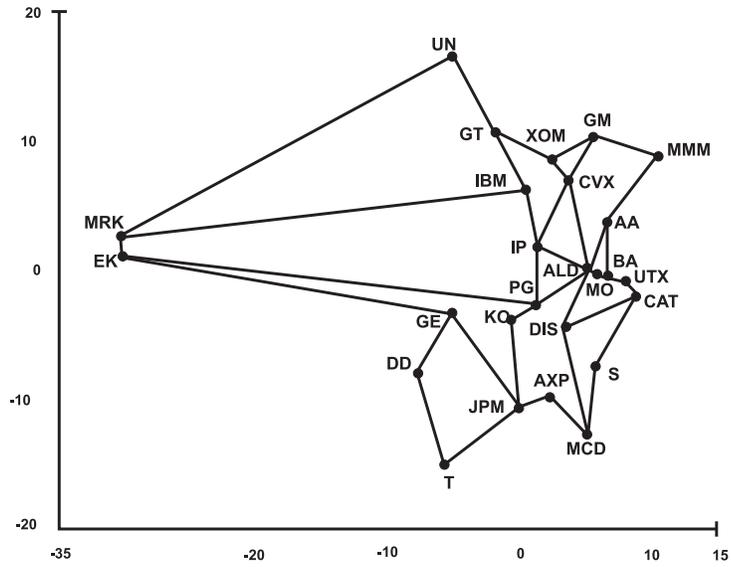}}
\caption{SOM-grid. Data taken for the year 2003.} \label{111}
\end{figure}

\begin{figure}[tbph]
\begin{center}
\epsfxsize=2.3in {\epsfbox[170 150 500 600]{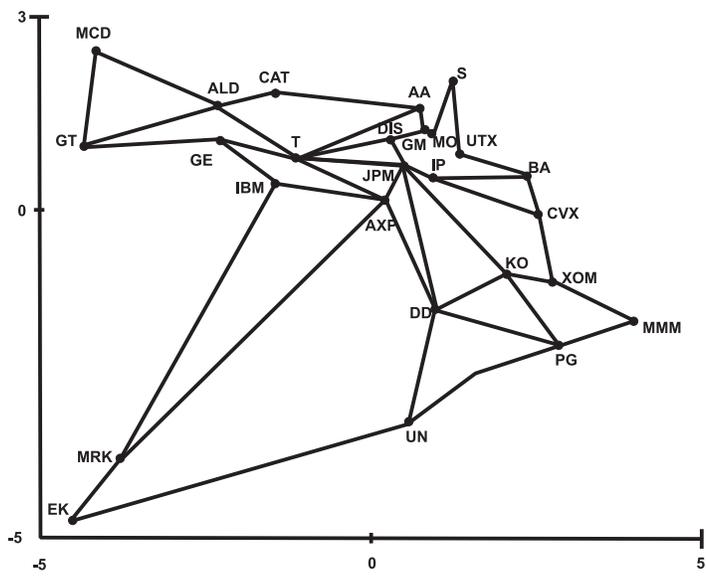}}
\end{center}
\caption{SOM-grid. Data taken for the years 2002-2003.} \label{14}
\end{figure}

Nevertheless, SOM - maps preserve their topological similarity as
a whole. This evidences close internal relationships between the
companies that determine the economy of the country. SOM - grid
for 2002 - 2003 (N= 490 working days) also proves this fact (Fig
4). Figures 1, 3 and 4 represent the dynamics of stock prices of
the considered companies for the  years 2002, 2003 and 2002-2003,
respectively. All the figures show that EK and MRK companies
maintain their position relative to the others and this proves
the hypothesis about their close relation (EK develops chemical
technologies in the area of photo, video films and polymeric
materials while MRK is a mining company). One can assume that MRK
is connected with production of raw materials for EK. Comparing
the figures it is easy to see that high technological companies
involved in developing the defensive technologies (for example,
BA, CAT, UTX) fall in the same cluster. Probably these
cooperative behavior is influenced by the events in Iraq in 2003.
In this case the stock prices behavior of these companies is very
similar. It should be noted that ALD company working in the area
of financial services and insurance is close enough to the
companies described above (BA, CAT, UTX). This fact can be
explained by the increased interest of people and businesses in
real estate and safety insurance.

\section{The GMDH analysis}

Dynamics of collective stock market behavior can be analyzed by
methods of multiple regression \cite{2}. But these methods take
into account the whole set of input data and overload the final
model (equation of regression). Self-organizing algorithms do not
have this disadvantage. Group Method of Data Handling (GMDH)
creates the model that includes only the most influential
variables \cite{3,7}. The GMDH algorithms are based on a
sorting-out procedure of model simulation and provide the best
model according to the criterion given by the researcher. This
model describes relations between their elements and the state of
the whole system. Most of GMDH algorithms use polynomial
referenced functions. General connections between input and
output variables can be shown by Volterra functional series. A
discrete analogue of Volterra series is Kolmogorov-Gabor
polynomial

\begin{equation}\label{polynom}
y=a_0+\sum_{i=1}^N a_i x_i +\sum_{i=1}^N \sum_{j=1}^N a_{ij} x_i
x_j + \sum_{i=1}^N\sum_{j=1}^N\sum_{k=1}^N a_{ijk} x_i x_j x_k +
...,
\end{equation}
where $y$ - output variable vector, $(x_1,x_2,...,x_N)$ - input
data, \\ $(a_1,...,a_N,...,a_{ij},...,a_{ijk},...)$ - vector of
coefficients or weights. Input data might consist of independent
variables, functional expressions or finite residues. The key
feature of GMDH algorithms is a partition of input data into two
subsets. The first one is used to compute coefficients of the
polynomial using the list square technique and to evaluate
internal error by some criterion. The second one is used to
calculate external error using information, which is not applied
for the coefficients computations. Principles of
self-organization manifest themselves in rationalization of
optimal polynomial search. Internal criterion monotonously
decreases when complexity of polynomials increases, simultaneously
external criterion passes its minimum. Then it is possible to
choose polynomial of optimal complexity, which is unique for this
criterion. In other words, we provide sorting-out procedure for
partial polynomials to find polynomial of optimal complexity
(optimal model). It shows the dependence of the output variable on
the most influential variables, which are chosen from all input
variables. External criterion reaches its minimum on optimal
model. Interpretation of the results is similar to multiple
regression logic: the bigger is the coefficient - the  more
influential is the variable near it. \par The combinatorial
algorithm was used for providing complete search between all
possible polynomials of the first order \cite{7}. The peculiarity
of this algorithm comprises gradual change of candidate
polynomials. In this case the number of variables gradually
increases from $1$ to $N$. For our investigation we use unbiased
criterion and variation criterion \cite{1}.
\par The complete set of polynomials includes $k=2^{27}-1$ models
and the computations require a lot of time. Thus, we take into
account the elements of the certain cluster and its neighbors
created by SOM algorithm (Fig.1). Therefore, it makes it possible
to analyze bigger set of elements relationships and obtain certain
hierarchy, to find the most influential variables in each
cluster. \par We estimated that for each cluster fewer a 15
vectors might play a role in stock behavior. As a rule, the
obtained polynomials of different complexity correspond to the
results presented by SOM. The number of the dependent variables
in the models varies from 3 to 8. In some cases GMDH models
describe relationships among elements of one cluster. For example,
relationships among companies S, MCD and MO are not evident
beforehand. Analytical expression for such relationships is

\begin{equation} Y_{MCD}= - 0.03 + 0.69X_S +
0.34X_{MO}.\end{equation} Considering relationships between
elements of the cluster I, which was described above, one obtains
the following model
\begin{equation}Y_{CVX}= - 0.029 + 0.7X_{GM} + 0.43X_{BA}.\end{equation}
It is easy to see that the model includes only three elements of
the cluster. It was not evident that relationships between these
three companies don`t depend on the stock behavior of the other
companies drawing up Dow Jones index. In the most cases the
relationships we obtain have bigger numbers of elements, some of
which have large coefficients. \par It should be noted that the
relationships between companies obtained by this method are not
reciprocal. Being bounded by the equation some companies more
sufficiently influence the stock price dynamics of others than
those of their own. For example, we consider cluster III, which
includes UN, MMM and PG (Fig.2). The obtained models are
\begin{equation}\label{*} Y_{PG}= 0.29 + 0.37X_{MMM} + 0.27X_{UN}
- 0.12X_{AXP} + 0.1X_T
\end{equation} and \begin{equation}\label{**}
Y_{MMM}= 0.07 + 0.64X_{UN} + 0.4X_{PG} - 0.13X_{KO} - 0.05X_{MO}.
\end{equation} These equations show that relationships between
elements of cluster III are essential. In particular, the dynamics
of $Y_{PG}$ depends on the dynamics of $X_{MMM}$ and vice versa.
In the same time $Y_{PG}$ also depends on the dynamics of
$X_{AXP}$ and $X_T$, but $Y_{MMM}$ depends on the dynamics of
other indices, such as $X_{KO},\; X_{MO}$.

\par We found that the relationships between prices dynamics of the companies are very robust.
Let us consider an example describing dynamics of cluster I.
\begin{equation}\label{exam} Y_{AA}= 0.02 + 0.57X_{GT} +
0.48X_{CAT} - 0.01X_{GM}.
\end{equation}
Easy to see that $X_{GT}$ is
the most influential element for $Y_{AA}$. The price $X_{CAT}$ has
less weight than other prices. Eliminating from consideration
$X_{GT}$ and leaving other elements we obtain a new model, which
has a new set of elements. But $X_{CAT}$ belongs to both models
and has the biggest weight in a new expression.
\begin{equation}\label{ex} Y_{AA}= - 0.004 + 0.46X_{CAT} +
0.35X_{CVX} + 0.28X_{GE}.
\end{equation}

The same procedure could be applied to the data for the year 2003
and makes it possible to find analytical expressions for the stock
market behavior of any company drawing up Dow Jones index. We do
not display these results in order to restrict the volume of the
paper. In addition, this investigation has a particular interest
for people dealing with specific finance investigation.
\section{Conclusions}

The results obtained by SOM algorithm are sufficiently
demonstrative and make it possible to understand deep
relationships inherent to such a complex system like stock market.
The GMDH algorithm, in its turn, makes it possible to establish
analytical dependence of the stock prices of the companies, which
have correlated behavior. The introduced self-organizing methods
complement each other. The obtained results are self-consistent
and allow to find an optimal non-overloaded model, which is easy
for economic interpretation. Using the combination  of such
methods promises to be very perspective.

\end{document}